\documentclass[runninghead]{llncs}
\usepackage[T1]{fontenc}

\usepackage{microtype}
\usepackage{xstring}
\usepackage{booktabs}
\usepackage{color}
\usepackage{array} 
\usepackage{tikz}
\usepackage{subfiles}
\usepackage{multicol}
\usepackage{multirow}
\usepackage{subfigure}
\usepackage{caption}
\usepackage{subcaption}
\usepackage{graphicx}
\usepackage{enumitem}
\usepackage{amsmath}
\usepackage[table]{xcolor}
\usepackage[dvipsnames]{xcolor}
\usepackage{hyperref}
\usepackage{xcolor, soul,todonotes} 
\definecolor{kmycolor}{rgb}{0.858, 0.188, 0.478}

\usepackage{graphicx}
\begin{document}
\title{Improving Conversational Recommendation with Contextual Adaptation of External Recommenders and LLM-based Reranking}

\author{Chuang Li\inst{1}\thanks{Work was done during a remote internship at SMU.}\orcidID{0009-0006-8112-3505}\and
Weida Liang\inst{1} \and
Hengchang Hu\inst{1}\and\\
See-Kiong Ng\inst{1}\and
Min-Yen Kan\inst{1}\and
Haizhou Li\inst{1,3}\and
Yang Deng\inst{1,2} 
}

\authorrunning{Li et al.}

\institute{National University of Singapore, Singapore \\
\and Singapore Management University, Singapore\\
\and Chinese University of Hong Kong, Shenzhen\\
\email{\{lichuang, weida\_liang, hengchanghu\}@u.nus.edu}\\
\email{\{seekiong, kanmy, haizhou.li, ydeng\}@nus.edu.sg}
}

\maketitle              

\begin{abstract}
We tackle the challenge of integrating large language models (LLMs) with external recommender systems to enhance domain expertise in conversational recommendation (CRS). Current LLM-based CRS approaches primarily rely on zero/few-shot methods for generating item recommendations based on user queries, but this method faces two significant challenges: (1) without domain-specific adaptation, LLMs frequently recommend items not in the target item space, resulting in low recommendation accuracy;
and (2) LLMs largely rely on dialogue context for content-based recommendations, neglecting the collaborative relationships among item sequences.
To address these limitations, we introduce the \textit{CARE (Contextual Adaptation of Recommenders)} 
framework. CARE (a) integrates external recommender systems as domain experts, producing candidate items through entity-level insights, and (b) customizes LLMs as rerankers to enhance the accuracy by leveraging contextual information.
Our results demonstrate that incorporating CARE framework significantly enhances recommendation accuracy of LLMs by an average of $54\%$ and $25\%$ for ReDial and INSPIRED datasets. The most effective CARE strategy involves LLMs selecting and reranking candidate items that external recommenders provide based on contextual insights. 


\end{abstract}

\section{Introduction}

A conversational recommender system (CRS) helps users achieve recommendation-related goals through multiple rounds of conversation~\cite{A-survey-jannach,li2023conversation,pramod2022conversationalsurvey}. Compared with traditional recommendation systems that mainly use entity-level information (\textit{e.g., item name, ID or attributes}) as input, CRS adopts conversational data (\textit{e.g., dialogue history}) as input, fully aligning to real-world scenarios \cite{li2023conversation,he2023large}. Aside from the input difference, CRS aims to accomplish two tasks: 1) recommending items based on both the user's query or interests, and 2) generating high-quality conversational responses~\cite{li2023conversation,li_user-centric_2022}.

\begin{figure*}[!t]
    \setlength{\abovecaptionskip}{5pt}   
    \setlength{\belowcaptionskip}{0pt}
\begin{center}
\includegraphics[width=0.95\textwidth]{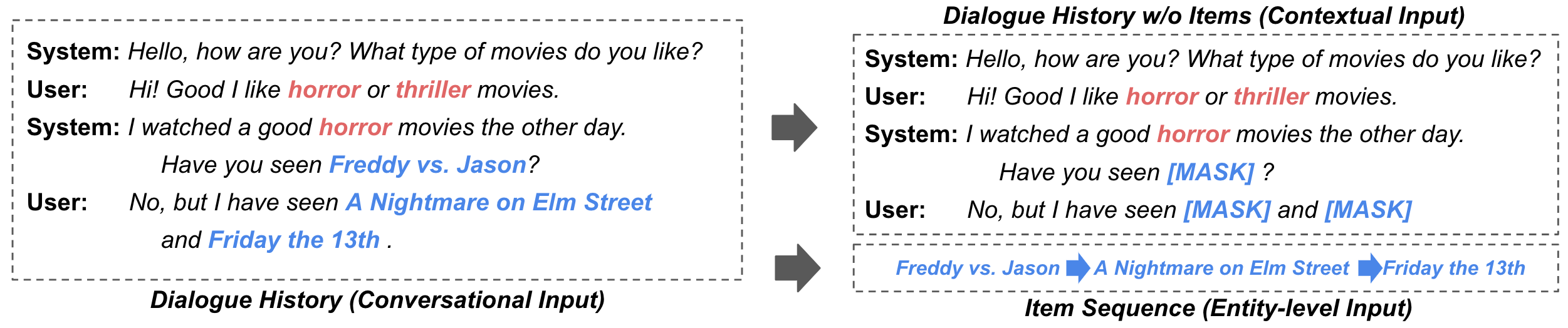}
\caption{{CRS input example (ReDial Dataset) with conversational, contextual and entity-level inputs. ({\color{blue}Blue}: item entities; {\color{red}Red}: attribute entities)}}
\label{example}
\vspace{-1.5em}
\end{center}
\end{figure*}

A key challenge for CRS is effectively leveraging dialogue history to generate accurate recommendations \cite{he2023large,EAR,li_towards_2018_ReDial,SAUR}. As illustrated in Figure~\ref{example}, CRS utilizes two main types of information. First, entity-level information extracted from the dialogue history, including attribute entities (e.g., ``user enjoys horror movies'') or item entities (e.g., ``historical movies watched by the user''). Second, contextual information expressed in natural language (e.g., ``user's opinion for a specific movie''). These two types of information are integrated throughout the conversation to guide the CRS in generating relevant recommendations.

With the advent of large language models (LLMs), studies have explored their effectiveness as conversational recommenders \cite{LLM_rec,he2023large,LLM_evaluate}. Leveraging their superior language capabilities, LLMs show impressive results in zero-shot CRS applications. However, two key challenges remain due to the limitations of 
{LLM-based CRS}: \textbf{\textit{1) item space discrepancy}} and \textbf{\textit{2) item information negligence}} {(\S~\ref{OOD})}.
The issue of \textit{item space discrepancy} arises when 
{LLMs} are pre-trained on data that differs from the domain-specific item space, causing LLMs to generate out-of-domain items, which fall outside the relevant target domain \cite{he2023large,xi2024memocrs}. 
The problem of \textit{item information negligence} stems from the zero-/few-shot learning setting, where LLMs lack access to item-specific training data \cite{ReFICR,chat-rec}. Furthermore, they lack comprehensive user histories for items and attributes, leading them to provide recommendations in cold-start scenarios predominantly. 
Consequently, these models cannot leverage collaborative patterns among users and items.
Instead, they focus on content-based queries from the dialogue history and rely heavily on contextual information in the dialogue history to make recommendations \cite{he2023large,DSI,he2024reindex}. 
These two limitations have been acknowledged in prior studies (e.g., \cite{chatcrs,CRS-agent:2023recommender,he2023large}), but remain unaddressed in subsequent CRS research. 

Inspired by advances in traditional recommender systems that integrate LLMs with external collaborative filtering techniques \cite{CF_challenge,CF_challenge2}, we propose the \textit{Contextual Adaptation of RecommEnders (CARE)} framework. 
CARE bridges LLMs and external entity-level recommenders via contextual adaptation, using item entities as intermediaries. 
Specifically, we first train a lightweight domain-specific recommender based on entity sequences in dialogue history \cite{zou_improving_2022,improving-sequential}. This recommender acts as a entity-level domain expert, providing insights in target domain (challenge 1). However, operating solely at the entity level limits its ability to capture user intentions.
For example, although both statements—\textit{``I like horror movies for tonight''} and \textit{``I like thrillers but I don’t want to watch tonight!''}—express a taste for scary films, they reflect opposite intentions in viewing options.
Therefore, we then adapt the recommender’s insight but enable LLMs as rerankers to engage contextually by selecting and reranking the entity-based candidates using dialogue history (challenge 2). Specially, we study the rules and scales of the LLM-based rerankers as different strategies defined in \S~\ref{prompt_llm}.
Our main contributions are summarised:
\begin{itemize}[leftmargin = *] 
\item We target two identified but unsolved challenges in LLM-based CRS: \textit{item space discrepancy} and \textit{item information negligence}. 
\item We introduce the CARE framework, using LLM-based rerankers with external recommenders to leverage both entity-level and contextual information for conversational recommendation, 
demonstrating different strategies to enhance the quality and control the freedom of LLM-based rerankers.
\end{itemize}

\section{Related Work} \label{related}

\textbf{Conversational Recommendation System (CRS)} aims to model user preferences and generate recommendations through multi-round of conversations \cite{A-survey-jannach,gao_advances_2021,li2023conversation}. The current research focus has shifted from the \textit{attribute-based approaches}, where the system and users exchange items or attribute entities using a template \cite{SAUR,EAR}, into \textit{conversational approaches}, where the system interacts with users through natural language \cite{li2018conversational,uniCRS,baseline_TPNet}. The majority of CRS in conversational approaches use language models (LMs; e.g., \textit{DialoGPT}) 
for learning-based preference modelling \cite{li_towards_2018_ReDial,INSPIRED-shirley,liu-etal-2021-durecdial2}. Particularly, LMs are trained to generate recommendations or system responses using word or entity embeddings encoded from the conversations or external knowledge graphs \cite{li_towards_2018_ReDial,uniCRS1,RecInDial,zhou_improving_2020}.

\textbf{Large Language Models for CRS.} LLMs have shown promising performance in CRS as zero-/few-shot conversational recommenders using item-based \cite{LLM_evaluate,LLM_rec} or conversational inputs \cite{he2023large,LLM_competitive_zero-shot,LLM-evaluation} to generate recommendation results. However, the performance of LLMs largely depends on their internal knowledge and their capability will notably diminish in domains with scarce knowledge \cite{chatcrs,LLM-evaluation}. Thus, LLMs are integrated with external agents \cite{LLM-expert,LLM_e-commerce,CRS-agent:2023recommender} to either directly provide the necessary knowledge resources or retrieve the essential knowledge resources from external knowledge base to improve their performance in domain-specific CRS tasks \cite{chatcrs,xi2024memocrs}. However, there is limited research in integrating traditional recommender systems to improve the item-specific capability for conversational recommendation tasks \cite{he2024reindex,improving-sequential}.

\textbf{Sequential Modelling in CRS.} Sequential modelling is well-suited for CRS applications for two reasons: (1) entities in CRS are mentioned in a sequential flow that leads to recommendations, and (2) cold-start scenarios in CRS often limit the collaborative filtering methods \cite{wang2019sequential,improving-sequential,li_towards_2018_ReDial,xi2024memocrs,improving-sequential}. Compared to RNN models, transformers have shown superior performance in sequential recommendation \cite{GRU4Rec,BERT4REC}. Motivated by the work in sequential recommendation \cite{BERT4REC}, the transformer-based sequential modelling framework is also applied to CRS by encompassing the entities mentioned in CRS conversations as a sequence using positional embedding \cite{improving-sequential} or knowledge-aware positional encoding \cite{UMAP}. However, the existing sequential modelling methods only use entity-level information while we introduce the framework to incorporate contextual information in modelling user preferences.

\begin{figure*}[!t]
    \setlength{\abovecaptionskip}{5pt}   
    \setlength{\belowcaptionskip}{0pt}
\begin{minipage}[t]{0.46\textwidth}
\includegraphics[width=\textwidth]{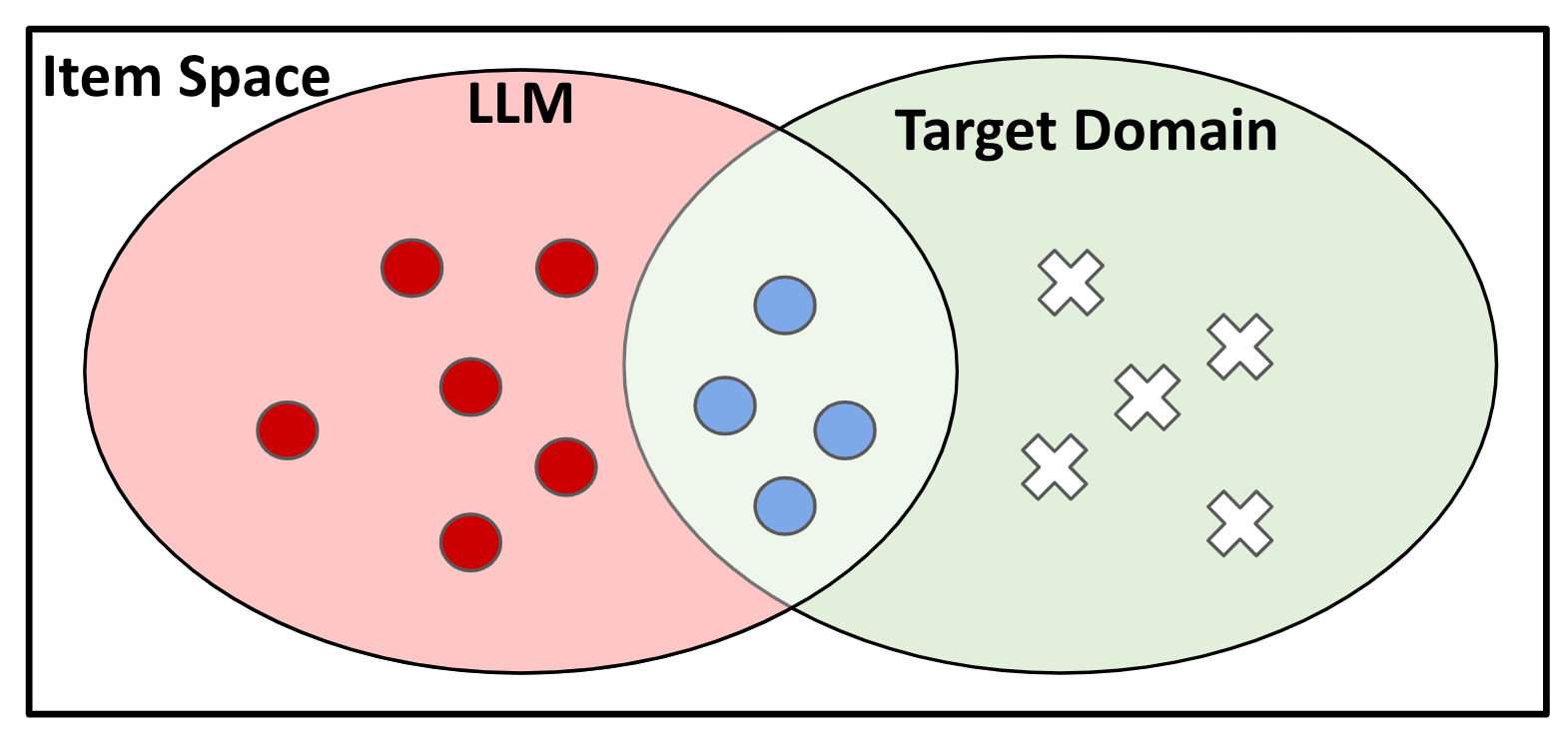}
\small
\caption{{Item space discrepancy between LLM and target domain. ({\color{blue}{Blue Dot}}: good recommendation results; {\color{red}Red Dot}: recommendations outside target domain; {\color{gray}White Cross}: items out of LLM's internal knowledge)}}

\label{p2}
\end{minipage}
\hfill
\begin{minipage}[t]{0.5\textwidth}
    \includegraphics[width= 0.9\textwidth]{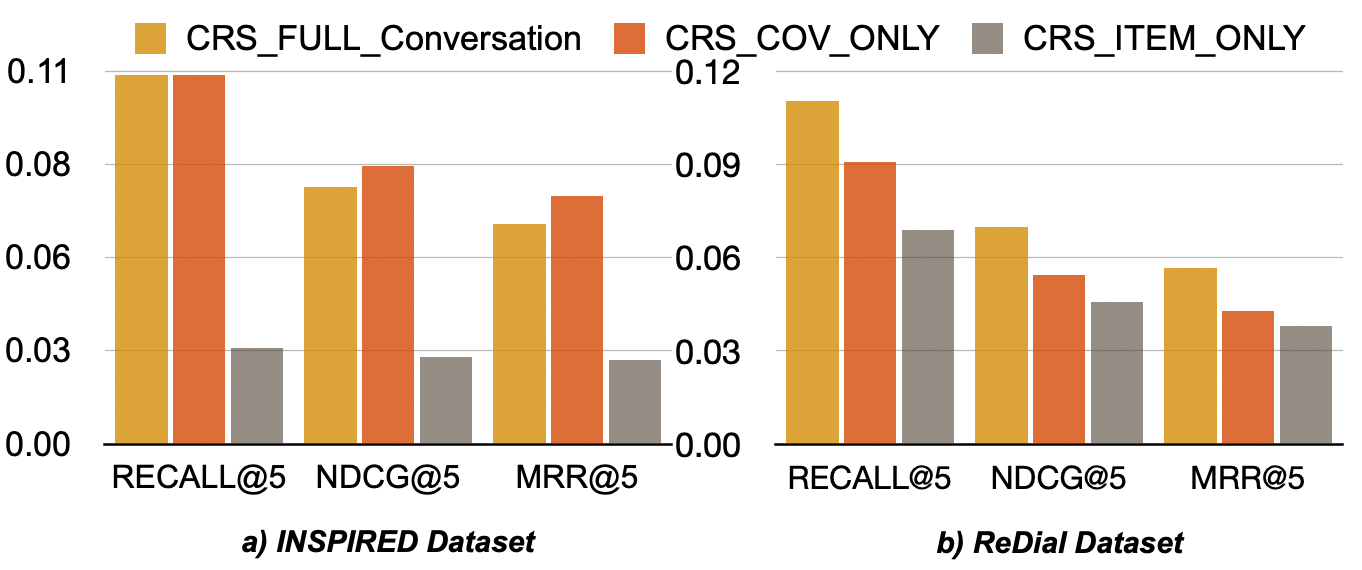}
    \caption{{Ablation study for LLM-based CRS with different levels (conversational or entity) inputs. ({\color{yellow}{Yellow}}: original conversational inputs in datasets; {\color{orange}{Orange}}: Contextual inputs w/o items; {\color{gray}Gray}: Entity-level inputs w/o conversations)}}
\label{p1}
\end{minipage}
\vspace{-1em}
\end{figure*}

\section{Preliminaries} \label{Preliminary} 
\textbf{Task Formulation.}
We consider the CRS scenario where a system $sys$ interacts with a user $u$. Each dialogue contains conversation turns denoted as $Cov$. The sequence of entities mentioned by $u$ or $sys$ is extracted as a historical entity sequence $Seq = [e_1, e_2, ..., e_n]$.
As this paper limits its scope to the \textit{recommendation task} in CRS, the target function for CRS is expressed as 
the generation of a recommendation list of items $i_1, i_2, ..., i_n$,
given the dialogue history, $Cov$ and sequence of historical entities $Seq$.
\begin{equation}
\label{LLM-func}
i_1, i_2, ... , i_n = CRS(Cov,~Seq)
\end{equation}

\subsection{Main Problems and Key Challenges}\label{OOD}

\paragraph{\textbf{Item Space Discrepancy.}}
We illustrate the issue of item space discrepancy in Figure~\ref{p2}. LLMs generate zero-shot recommendations based solely on their internal knowledge (left circle in red), which can significantly differ from the target item space (right circle in green) in CRS \cite{he2023large,chatcrs}. The overlap between these spaces represents where LLMs can make accurate recommendations (blue dots). The red dot indicates recommendations outside the target item space, while the white cross highlights target space items that the LLM cannot recommend due to lack of exposure or knowledge. 
We show the reduction of out-of-domain items from LLMs using our framework in Figure~\ref{OOD_figure}, which directly contributes to the performance improvement.
While recommending out-of-domain items may occasionally produce creative outputs, like AI-generated content (AIGC for recommendation) \cite{AIGC,evaluation-yoon2024evaluating}, in real-world services such as e-commerce, these recommendations often misalign with the platform’s offerings, which can lead to user disappointment \cite{he2023large,improving-sequential,uniCRS1}. Based on this need, we apply policy in different strategies to control the freedom of LLMs in reranking (\S~\ref{strategies}).

\paragraph{\textbf{Item-specific Information Negligence.}} \label{single}
LLMs excel at mapping user descriptions or queries to target items \cite{he2023large,DSI}.
However, a key limitation is their lack of understanding of the item-specific information from the item entities mentioned by the users or system  \cite{he2023large,chatcrs}. 
To investigate this, we follow the approach in \cite{he2023large} and evaluate LLM performance by separating different input types with or without item entities (Figure~\ref{example}): (1) the full conversation (\textit{CRS\_FULL\_Conversation}), which includes the complete dialogue history; (2) the entity-level item sequence (\textit{CRS\_ITEM\_ONLY}), consisting solely of item entities ($Seq$); and (3) a context-only input (\textit{CRS\_COV\_ONLY}), where item entities were excluded from the dialogue history ($Cov - Seq$). The results in Figure~\ref{p1} demonstrate that LLMs primarily base their recommendations on contextual information, neglecting the item-specific information in the item entity sequence. While traditional recommenders are experts in entity-based information. To fill the gap, we propose CARE-CRS to integrate both contextual and item-specific information for conversational recommendation via LLM-based reranking.

\begin{figure*}[t!]
    \centering
    \setlength{\abovecaptionskip}{5pt}   
    \setlength{\belowcaptionskip}{0pt}
    \begin{center}
    \includegraphics[width=0.95\textwidth]{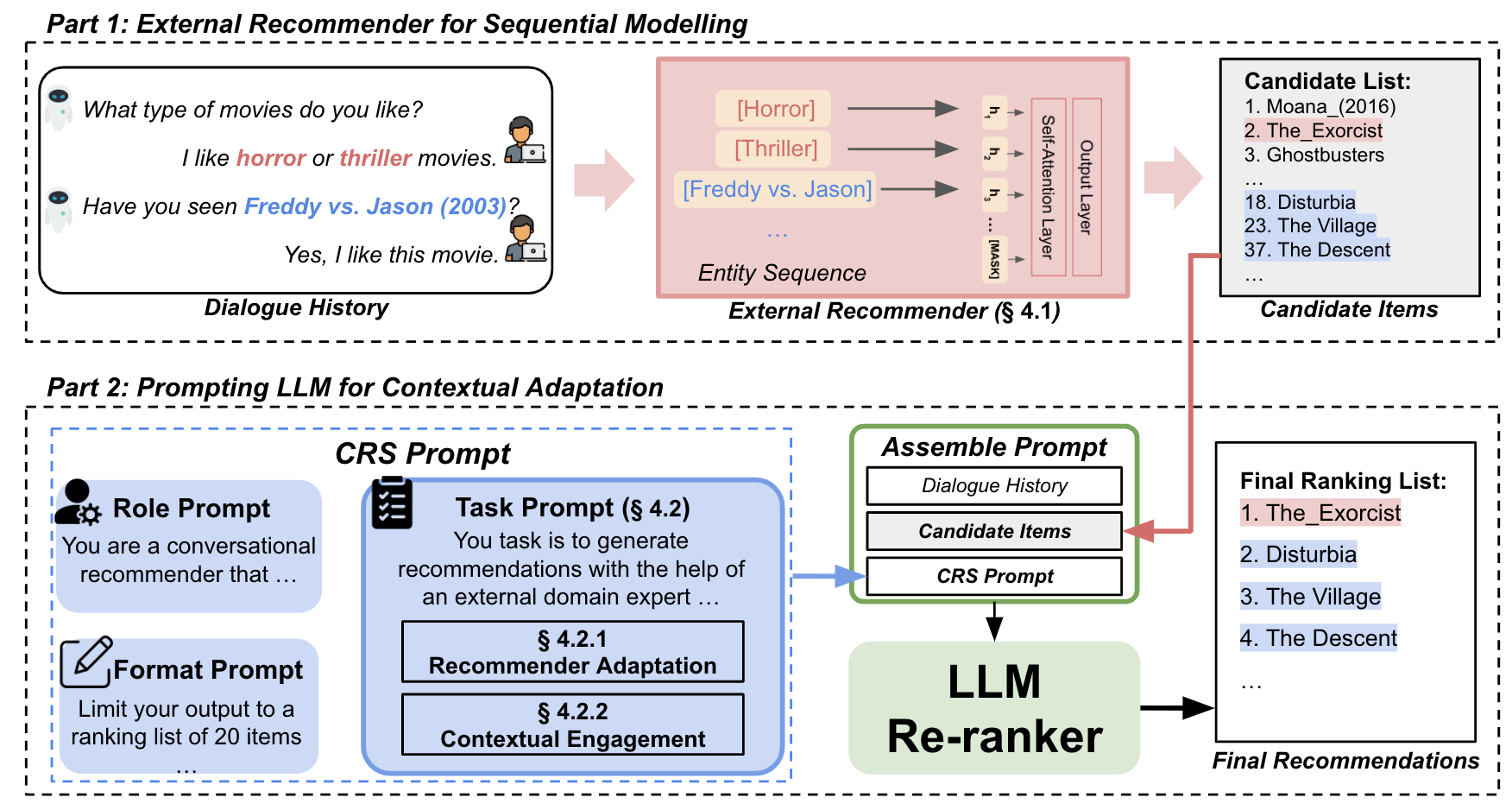}
    
    \caption{\textbf{CARE Framework}: Conversational inputs are firstly passed to an external recommender for entity-level sequential modelling (\S~\ref{seq_model}). The candidate items output and original dialogue history are jointly used to prompt LLMs for the contextual adaptation and generate conversational recommendations (\S~\ref{prompt_llm}).} 
    \vspace{-2em}
    \label{overall}
    \end{center}
\end{figure*}

\section{CARE-CRS} \label{Methods}

The CARE-CRS framework (Figure~\ref{overall}) consists of two parts: 1) external recommenders processes entities in dialogue history for sequential modelling and 2) LLMs as rerankers for conversational recommendations.

\subsection{Transformer-based Recommender for Sequential Modelling}\label{seq_model}

The external recommender system takes entity sequences as input. To generate this sequence, we first apply entity extraction scripts to align the entities in the dialogue with the ones in knowledge graphs  (e.g., movies, actors, attributes), as shown in Figure~\ref{example} \cite{DBpedia,improving-sequential,uniCRS1}. The script includes 3 main steps: 1) the full dialogue is tokenized into short entity-level terms, 2) the similarity of each entity in dialogue and knowledge graphs are compared, and the final entities are selected based on the similarity score and 3) the selected entities are converted as a sequential list for each dialogue. For example, in dialogue 
``User: I want to watch a movie with Tom Cruise; System: How about watching Mission Impossible?'', the corresponding entity sequences is [`Tom Cruise', `Mission Impossible']. 
These step converts the dialogues into entity-level  inputs for the sequential modelling.

While the external recommender can be implemented using any system with entity-level inputs and outputs, we adopt a transformer architecture for its superior performance in sequential modelling, as shown in prior works \cite{UMAP,improving-sequential,GRU4Rec,kang_self-attentive_2018}, Notably, we deliberately exclude LLMs from the recommender to preserve a clear modular separation from the LLM-based re-rankers. The model has three components: an embedding layer, a self-attention layer, and an output layer \cite{BERT,BERT4REC}.
The embedding layer computes each entity's embedding combing graph embedding $h_K$ and positional embedding $h_P$ for final embedding $h_i^0$ {as $h_K + h_P$}, as in \cite{UMAP,improving-sequential}. 
The self-attention layer uses a Multi-head Self-attention mechanism to learn the final sequential representation $H$, formulated as:
\begin{equation}
  H = \text{Self-Attention}(h_1^0, h_2^0, ..., h_n^0, \text{\textit{[MASK]}}) 
\end{equation}

The output layer projects the final hidden states of sequential representation $H$ to the item space $Out_k$ using two layers of neural networks, where $W_1$, $W_2$, $b_1$, and $b_2$ are the weights and biases of the feed-forward networks. The final outputs are converted into text format for LLMs.
\begin{equation}
  \text{Out}_k = [\text{out}_1 \dots, \text{out}_k];~\text{out}_i = \text{GELU}(H W_1^T + b_1) W_2^T + b_2
\end{equation}


\subsection{Prompting LLMs for Contextual Adaptation}\label{prompt_llm}

Without external recommenders, the existing approach uses instructional prompts to guide LLMs in generating recommendation outputs from dialogue history \cite{he2023large}. {As shown in Figure \ref{overall}} (Part 2), a prompt consists of three components:
1) \textbf{role prompt ($P_R$)} that defines the system’s role,
2) \textbf{format prompt ($P_F$)} that constrains the output format for future post-processing, and 
3) \textbf{task prompt ($P_T$)} that specifies the task and function of the system.
The final prompt, $P_{CRS} = P_R + P_T + P_F$, integrates these three components. The overall process is formulated as follows:
\begin{equation}
\label{LLM-func1}
i_1, i_2, ... , i_n = LLM(~P_{CRS}, Cov)= LLM(~(P_R  + P_F + P_T), Cov)
\end{equation} 
In our setup, to better integrate external recommenders, we update the task prompt $P_T$ from two key perspectives: 1) Recommender Adaptation and 2) Contextual Engagement. These updates enable the LLMs to effectively leverage contextual information, facilitating their ability to generate, select, or rerank the entity-level recommendations derived from an external recommender system.

\subsubsection{Adaptation of Recommender as Domain Expert.}\label{recommender_adapt}
LLMs demonstrate significant potential when integrated with external sub-systems to form agent-based \cite{agent1,LLM_agent} or expert-guided systems \cite{LLM_expert}. Adaptation refers to the challenge of effectively introducing and describing these external systems so that LLMs can better understand and collaborate with them as domain experts \cite{wang2023adapting} (Recommender Adaptation in Figure~\ref{overall} Part 2). We design three methods to adapt the recommender systems as domain experts: 1) Direct Prompting, 2) Recommender Description, and 3) Self-Reflection (colored in orange). Each approach modifies the task prompt $P_T$ in the LLM to facilitate this adaptation.
\begin{itemize}[leftmargin = *]
    \item \textbf{Direct Prompting:} Direct introduce the recommender as a domain expert without any description of its function or input/output format.
\end{itemize}
     \begin{tikzpicture}
  \node[draw, rectangle, thick, inner sep=2pt, text width=12.5cm, align=left] {
     \\\textbf{Task Prompt ($P_T^{m=1}$):} \textit{\color{orange}{``To help you with the recommendation, we introduce a domain expert who provides recommendations based on the training data and you can use the domain expert's recommendations as examples for your output''}}
    };
\end{tikzpicture}
\begin{itemize}[leftmargin = *]
    \item \textbf{Description of Recommender:} Briefly describe how the recommender works to model the sequential inputs and generate the candidate items. 
\end{itemize}
     \begin{tikzpicture}
  \node[draw, rectangle, thick, inner sep=2pt, text width=12.5cm, align=left] {
     \\\textbf{Task Prompt ($P_T^{m=2}$):} \textit{\color{orange}{``To help you with the recommendation, we introduce a domain-expert which is a recommender for sequential modelling that uses the entities mentioned in the dialogues to generate a ranking list of items.''} }
    };
\end{tikzpicture}
\begin{itemize}[leftmargin = *]
    \item \textbf{Self Reflection:} Step by step lead the LLM to examine all the resources of the recommender (e.g., code, paper or data samples). After each resource, we ask LLM to self-reflect, correcting or generating a new prompt to describe the recommender \cite{selfdiscover}. We repeat these steps until the prompts generated from the LLMs has minor or no difference with the previous one and LLMs confirm its confidence about its output. We then fix it as the final task prompt $P_T^{m=3}$ for self-reflection.
\end{itemize}
     \begin{tikzpicture}
  \node[draw, rectangle, thick, inner sep=2pt, text width=12.5cm, align=left] {
     \\\textbf{Task Prompt ($P_T^{m=3}$):} \textit{\color{orange}{``To help you with the recommendation, you can access an advanced recommendation system that specializes in enhancing conversational recommendations by leveraging both the sequence of entities mentioned in a conversation and external knowledge embedded in knowledge graphs. This system ...''}} 
    };
\end{tikzpicture}

\subsubsection{Contextual Engagement for Conversational Recommendations.}\label{contextual_engag}
After introducing the recommender as a domain expert, we propose contextual engagement strategies that enable LLMs to learn item space information from the expert and incorporate contextual data to refine its recommendations \cite{he2023large,reranking}. This allows LLMs to generate target domain-specific recommendations and correct external recommender errors, such as \textit{1) context-agnostic rankings, where the recommender ignores the explicit preference over an item in the context} or \textit{2) popularity bias, where the statistically popular item is over-ranked}, both are common issues in learning-based approaches  \cite{improving-sequential,pop_bias}. In addition, as practiced in major industrial reranking systems, we design policy to control the freedom of LLM generation (colored in blue) \cite{rerank3,rerank1,rerank2}.
We introduce three contextual engagement strategies: 1) reranking and 2) selection-then-reranking, modifying prompt $P_T$ and formulated in the format of [{\color{orange}{Adaptation}} + {Engagement} + {\color{blue}{Policy}}].

\begin{itemize}[leftmargin = *]  
    \item \textbf{Reranking:} Provide the same number of recommendations as the desired output and ask the model to rerank the candidate items using dialogue history. 
\end{itemize}
\begin{tikzpicture}
  \node[draw, rectangle, thick, inner sep=2pt, text width=12.5cm, align=left] {
     \\\textbf{Task Prompt ($P_T^{s=1}$):} \textit{{\color{orange}{To help ... }} You need to \textbf{rerank} the recommendations, placing the domain expert's suggestions in the appropriate order based on your understanding of the dialogue history. \color{blue}{You \textbf{cannot} generate items beyond ...} }
  };
\end{tikzpicture}
\begin{itemize}[leftmargin = *]  
    \item \textbf{Selection-then-Reranking:} Show a larger set of recommendations and ask the model to select from them, rerank the selected items, to form a ranked list. 
\end{itemize}
\begin{tikzpicture}
  \node[draw, rectangle, thick, inner sep=2pt, text width=12.5cm, align=left] {
     \\\textbf{Task Prompt ($P_T^{s=2}$):} \textit{{\color{orange}{To help ...}} You need to \textbf{select} the most appropriate items from the domain expert's recommendations and \textbf{rerank} them in a ranked order based on dialogue history. \color{blue}{You \textbf{cannot} generate items beyond ...} }
  };
\end{tikzpicture}

Given the dialogue history $Cov$ and the top-K candidate items from the external recommender $Out_k$ as inputs, LLM is prompted to generate $n$ outcomes with adaptation methods $m$, contextual engagement strategies $s$, using the final prompt $P_{CARE}=(P_R + (P_T^m + P_T^s) + P_F)$ as formulated in:
\begin{equation}
\label{4.3}
\begin{split}
i_1, i_2, ... , i_n &= LLM(~P_{CARE}, Cov, Out_k)\\
                    &= LLM(~(P_R + (P_T^m + P_T^s) + P_F), Cov, Out_k)
\end{split}
\end{equation}

    


    

\begin{table}[t]
    \setlength{\abovecaptionskip}{6pt}   
    \setlength{\belowcaptionskip}{0pt}
\small
    \centering
    \begin{tabular}{lccccc}
        \toprule
        Dataset & \#Dialogues & \#Turns & \#Users & \#Items & Avg. \#Entities/Dialogue  \\
        \midrule
        \textbf{\textit{ReDial}} & 	$11,348$ & $139,557$& $764$ & $6,281$ & $ 7.24$ \\
        \textbf{\textit{INSPIRED}} & $999$ & $35,686$& $999$ & $1,967$&$12.88$ \\
        \bottomrule
    \end{tabular}
    \caption{Statistics of datasets.}
    \label{data}
    \vspace{-0.5em}
\end{table}

\section{Experiments} \label{exp}

We address the following research questions (RQs) in this section: 
\begin{itemize}[leftmargin=*]
    \item \textbf{RQ1}: How is the performance of CARE-CRS in recommendation tasks? 
\item \textbf{RQ2}: How are contextual adaptation strategies optimised in CARE-CRS?
\item \textbf{RQ3}: How efficient is CARE-CRS in solving identified challenges and biases?
\end{itemize}

\subsection{Setup}
\textbf{Datasets \& Evaluation Metrics.}
The experiment is conducted on two public CRS datasets: ReDial \cite{li_towards_2018_ReDial} and INSPIRED \cite{INSPIRED-shirley}.
The example of the dataset is shown in Figure~\ref{example} with statistics shown in Table~\ref{data} and we follow the original data split (8:1:1) \cite{crslab}.
We evaluate our model and baselines mostly on their recommendation abilities using metrics HIT@K, MRR@K and NDCG@K with K in $\{5, 10\}$, and response evaluation is not considered in this work \cite{uniCRS,he2023large,li_towards_2018_ReDial,RecInDial,uniCRS1}. 

\noindent\textbf{Baselines.} To compare our model with the existing CRS works, we adopt the baselines including 3 types of models. a) Benchmarks(BMK) released with datasets:
\textbf{ReDial} \cite{li_towards_2018_ReDial} use auto-encoder and \textbf{INSPIRED} \cite{INSPIRED-shirley} fine-tunes transformer-based LMs for recommendation; b) Learning-based approaches: \textbf{KBRD} \cite{chen_towards_2019} adopts external knowledge graph from DBpedia \cite{DBpedia}, \textbf{KGSF} \cite{zhou_improving_2020} use semantic fusion to align the representation of entities and dialogue history, \textbf{SASRec} \cite{kang_self-attentive_2018} use self-attention structure and \textbf{UniCRS} \cite{uniCRS} use prompt learning as a unified approach to jointly improve recommendation and response genenration; c) LLM-based methods: \textbf{ZSCRS} \cite{he2023large} studies the zero-shot conversational recommendation capability of LLM (ChatGPT), \textbf{Llama 3} \cite{grattafiori2024llama3herdmodels} implements ZSCRS using open-sourced LLM (Llama3-8B-Instruct) and \textbf{MemoCRS} \cite{xi2024memocrs} retrieve memory-based and collaborative information from the user's profile for preference modelling and generation (GPT4).

\noindent\textbf{Implementation Details.} 
We adopt the data pre-processed from the open-source CRS toolkit CRSLab \cite{crslab} and use closed-sourced ChatGPT and open-sourced Llama 3 \cite{grattafiori2024llama3herdmodels}\footnote{ChatGPT: gpt-3.5-turbo-0125; Llama 3: meta-llama/Meta-Llama-3-8B-Instruct}.
The temperature of ChatGPT is set as 0 to ensure the same output with fixed input tokens.  For the sequential modelling recommender, we follow the existing settings \cite{UMAP,improving-sequential} using 2 layers and 2 attention heads with an SGD optimizer and learning rate of 5e-3, dropout rates of 0.2. 
For our baselines, we replicate UniCRS \cite{uniCRS1} using their source code and other models using open-sourced CRSLab \cite{crslab}. For MemoCRS \cite{xi2024memocrs},  as their memory resources are not published, we report the results from the paper.

\begingroup
\setlength{\tabcolsep}{1.5pt} 
\renewcommand{\arraystretch}{1.2} 
\definecolor{darkgray}{rgb}{.75,.75,.75}
\definecolor{lightgray}{rgb}{.9,.9,.9}
\begin{table*}[!t]

\scriptsize
\centering
\fontsize{8pt}{8pt}\selectfont
\begin{tabular}{@{}c|l|ccc|ccc@{}}
\toprule

\multicolumn{2}{c}{\multirow{3}{*}{\textbf{Models}}}&\multicolumn{3}{c}{\textbf{ReDial}}& \multicolumn{3}{c}{\textbf{INSPIRED}}\\\cmidrule(lr){3-5} \cmidrule(lr){6-8} 
\multicolumn{2}{c}{}  & H@5/10            & M@5/10           & N@5/10        & H@5/10          & M@5/10            & N@5/10          \\ \midrule

\multirow{2}{*}{\rotatebox[origin=c]{90}{{BMK}}} & \textbf{ReDial}  & .029/.041      & .017/.019       & .020/.024  & .003/.003      & .001/.001        & .002/.002     \\
&\textbf{INSPIRED} & .099/.168 & .050/.059&.062/.084 &.058/.097 & .041/.046 & .046/.058  \\ \midrule
\multirow{4}{*}{\rotatebox[origin=c]{90}{{Learning}}}&\textbf{KBRD}              & .082/.140     & .034/.042       & .046/.065   & .019/.032      & .007/.009        & .010/.014       \\
&\textbf{KGSF}               & .091/.138     & .039/.045       & .051/.066    &   .016/.029   & .006/.007        & .008/.012      \\

&\textbf{SASRec}&    .036/.056& .020/.023& .024/.030& .088/.149& .054/.062& .062/.082\\
&\textbf{UniCRS}               & .101/.161    & .039/.047      & .054/.073     & .091/.106      & .062/.064& .070/.074         \\\midrule
\multirow{3}{*}{{\rotatebox[origin=c]{90}{{LLM}}}}& \cellcolor{lightgray}\textbf{ZSCRS}               & .111/.165    & .057/.064      & .070/.087    & \cellcolor{lightgray} .109/.142     & \cellcolor{lightgray} {.065/.069}        & \cellcolor{lightgray} {.076/.086}    \\
& \cellcolor{lightgray}\textbf{MemoCRS}          & \cellcolor{lightgray}{.136/.215}    & \cellcolor{lightgray}{.072/.082}       & \cellcolor{lightgray}{.088/.113}   & NA      & NA        & NA    \\\cmidrule(lr){3-8} 
& \cellcolor{darkgray}\textbf{CARECRS}  & \cellcolor{darkgray}{.194*/.248*} &\cellcolor{darkgray}{.133*/.140*}           &\cellcolor{darkgray}{.148*/.166*} &\cellcolor{darkgray}{.144*/.169*} &\cellcolor{darkgray}{.090*/.093*} & \cellcolor{darkgray} {.103*/.111*} \\
 &  \color{red}{Improvement}  & \color{red}{42.7/15.4\%} 
 & \color{red}{84.7/70.7\%}
 & \color{red}{68.2/46.9\%}
 & \color{red}{22.0/11.9\%}
 & \color{red}{38.5/29.2\%}
 & \color{red}{32.1/16.8\%}
\\

\bottomrule
\end{tabular}
\caption{Results of different models on recommendation task in H/M/N (HIT/MRR/NDCG). Best results from \colorbox{darkgray}{our model} and \colorbox{lightgray}{baselines} are coloured; Symbol~* indicates statistical significance with $p < 0.05$.}
\vspace{-2.5em}
\label{result_overall}
\end{table*}
\endgroup
\subsection{Experimental Results (RQ1)} 

\vspace{-0.5em}

\begin{figure*}[!t]
    \centering
    \setlength{\abovecaptionskip}{5pt}   
    \setlength{\belowcaptionskip}{0pt}
    \begin{minipage}[t]{.41\textwidth}
        \centering
            \includegraphics[width=\textwidth]{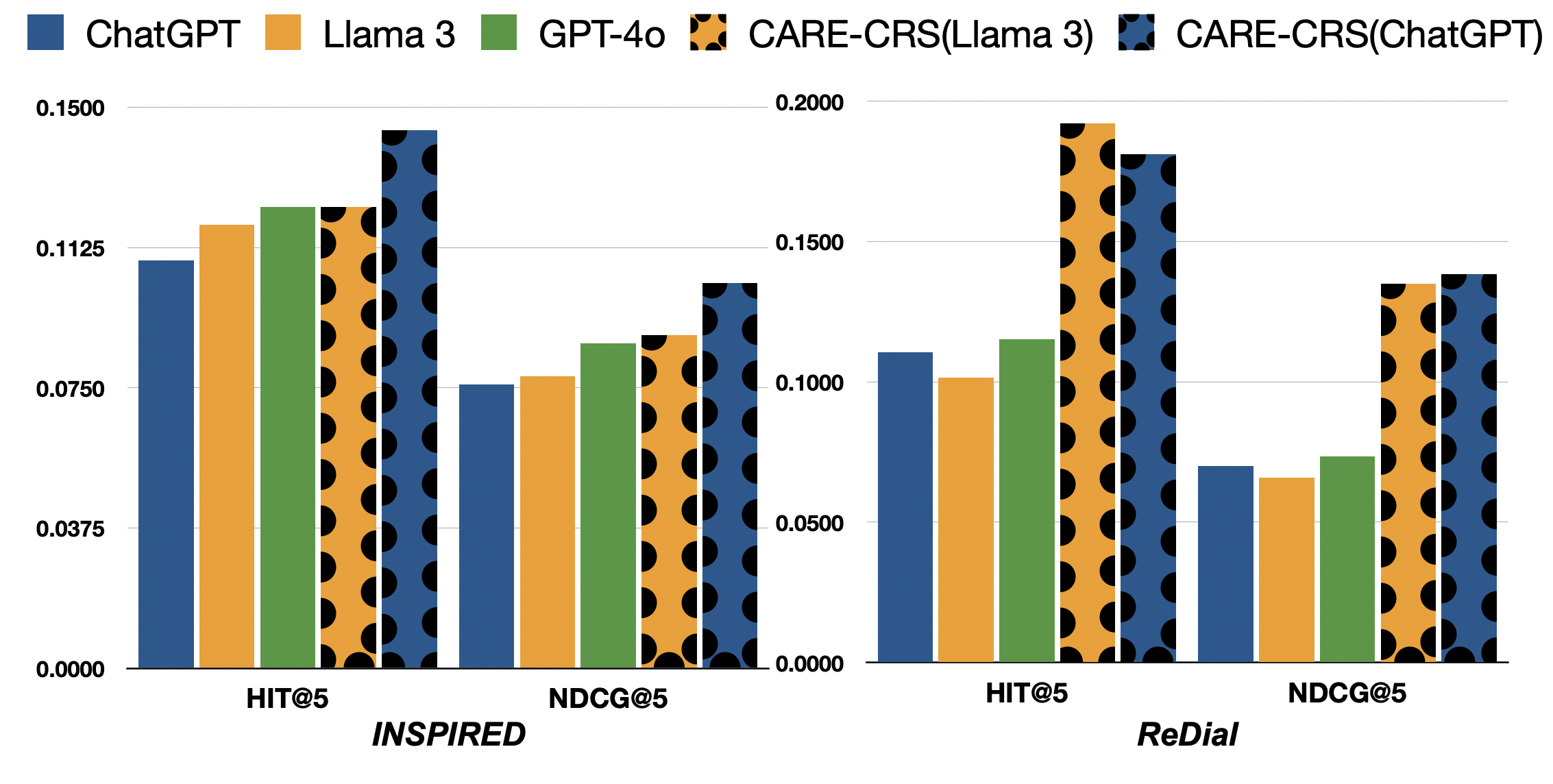}
        \caption{Comparison between open-sourced LLMs and {ChatGPT} in recommendation accuracy.}
        \label{LLM3}
    \end{minipage}
    \hfill 
    \begin{minipage}[t]{.57\textwidth}
        \centering
        \includegraphics[width=0.9\textwidth]{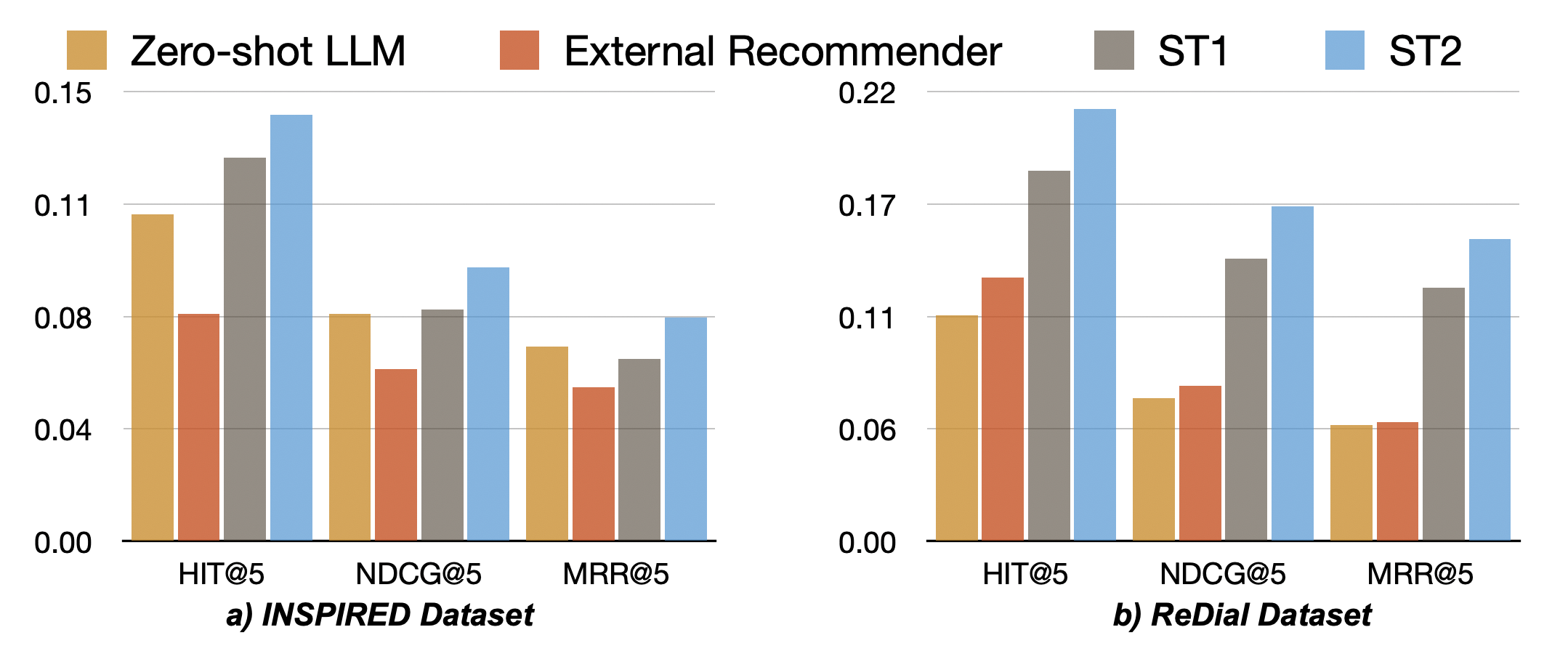}
        \caption{Analysis of different CARE strategies in CARE-CRS and ST1/ST2 stands for Reranking/Selection-then-Reranking.}
        \label{AB_strategy}
    \end{minipage}
\end{figure*}

\paragraph{\textbf{Recommendation Performance.}} Table~\ref{result_overall} presents the recommendation task results. We report the result of CARE-CRS using GPT3.5 as the base model and the optimized prompt with recommender description and selection-then-reranking as the best strategy (\S~\ref{strategies}).
All LLM-based approaches significantly outperform learning-based and benchmark baselines, which highlights the strong ability of LLMs to handle conversational queries for CRS applications \cite{he2023large,xi2024memocrs}. Our proposed CARE-CRS framework, which integrates contextual recommender adaptation, surpasses all state-of-the-art LLM-based methods with statistically significant improvements for top-5 and -10 accuracy across all metrics on both datasets. The performance gain for the top-5 metrics is generally higher than that for the top-10, with an average increase of $65\%$  and $31\%$ over the best LLM baseline.
In Figure~\ref{LLM3}, we show the recommendation results of CARE-CRS using open-sourced Llama 3 with 8B parameters. We adopt the GPT-4o \footnote{gpt-4o-08-06} models as a strong baseline to compare the results before and after applying CARE framework. For both datasets, the original recommendation performance of LLMs is lower than GPT-4o. However, after applying the CARE framework, they all outperform the GPT-4o in the recommendation performance, showing the efficiency and robustness of our framework for both the open- and closed-sourced LLMs. 

Compared to those leading LLM-based approaches, our method significantly enhances zero-shot performance by integrating a lightweight entity-level recommender for contextual adaptation, which benefit from a) sufficient candidate items, reducing the ratio of out-of-domain recommendations (discussed in \S~\ref{RQ3}) and b) entity-level sequential modelling from the recommender, incorporating both item-based and contextual information for final recommendations.


\begin{figure*}[!t]
    \setlength{\abovecaptionskip}{5pt}   
    \setlength{\belowcaptionskip}{0pt}
\begin{center}

\includegraphics[width=0.9\textwidth]{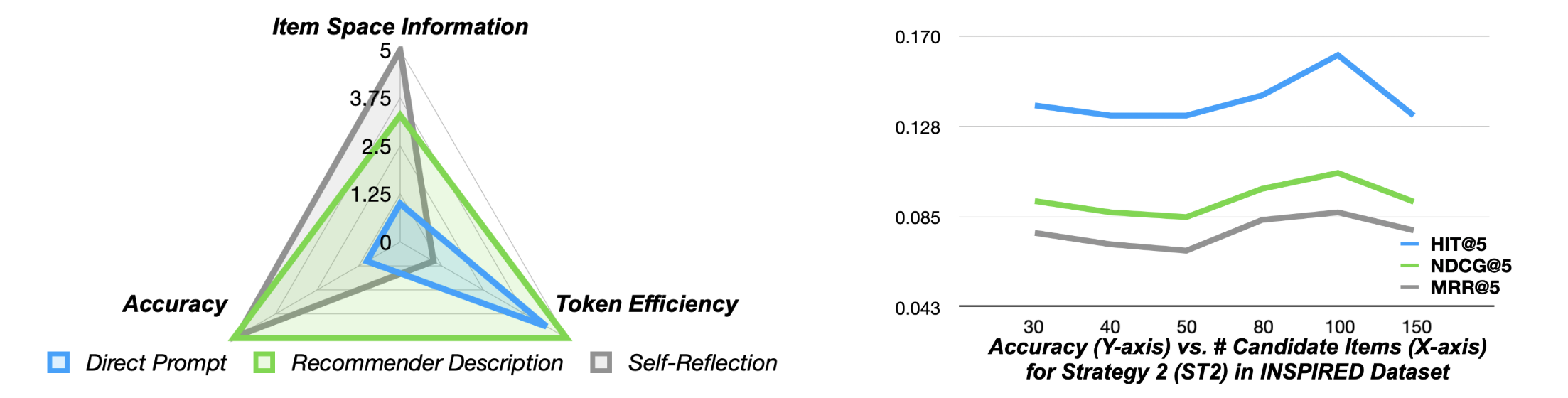}
\caption{Ablation study on adaptation methods(left) and candidate numbers(right).}
\label{AB_number}
\end{center}
\vspace{-3em}
\end{figure*}

\paragraph{\textbf{Training Efficiency and System Complexity.}} In contrast to learning-based baselines like UniCRS, which fully fine-tune a language model such as DialoGPT \cite{zhang2019dialogpt} with 762M parameters, our approach leverages a zero-shot LLM inference through API without fine-tuning. The only training involved is for the external recommender, which has 2 layers and 2 attention heads using a transformer structure, totalling 2.7 M parameters—less than 1\% of the size of the baseline models. As our overall framework is modular, both the external recommender and the LLMs can be replaced with newer models. Future work can further enhance training efficiency by using alternative pre-trained recommendation systems in the target domain, eliminating the need for additional training. 

\subsection{Discussion on CARE-CRS Methods and Strategies (RQ2)}\label{strategies}

We propose several methods for contextual adaptation between LLMs and recommenders in \S~\ref{Methods}. This section presents three ablation studies to evaluate and optimize the approaches for recommender adaptation, contextual engagement, and managing the number of candidate items.

\paragraph{\textbf{Contextual Engagement.}} \label{dis_contextual_engagement}
Figure~\ref{AB_strategy} illustrates the recommendation accuracy of the zero-shot LLM, external recommender, and the CARE framework using contextual engagement strategies (\textit{ST1}, \textit{ST2} corresponding to \textit{reranking}, and \textit{selection-then-reranking}). The performance of the zero-shot LLM and external recommender varies by dataset: the LLM excels in the INSPIRED dataset, while the external recommender performs better in ReDial. This discrepancy arises from the entity composition of each dataset (Table~\ref{data}). Because INSPIRED dataset contains more attribute entities, which provides rich context information for LLMs. This conclusion is further substantiated by Table~\ref{result_overall}, which shows that CARE-CRS, primarily introducing item-based information, achieves a significantly greater performance gain on ReDial than on INSPIRED. 
Despite dataset differences, \textit{ST1} and \textit{ST2} consistently outperform both the zero-shot LLM and external recommender by leveraging reranking or selection based on context. The consistency among the two strategies shows that the CARE framework is robust across different types of datasets, whether item- or attribute-rich. 


\begin{figure*}[!t]
    \setlength{\abovecaptionskip}{5pt}   
    \setlength{\belowcaptionskip}{0pt}
\begin{center}

\includegraphics[width=0.9\textwidth]{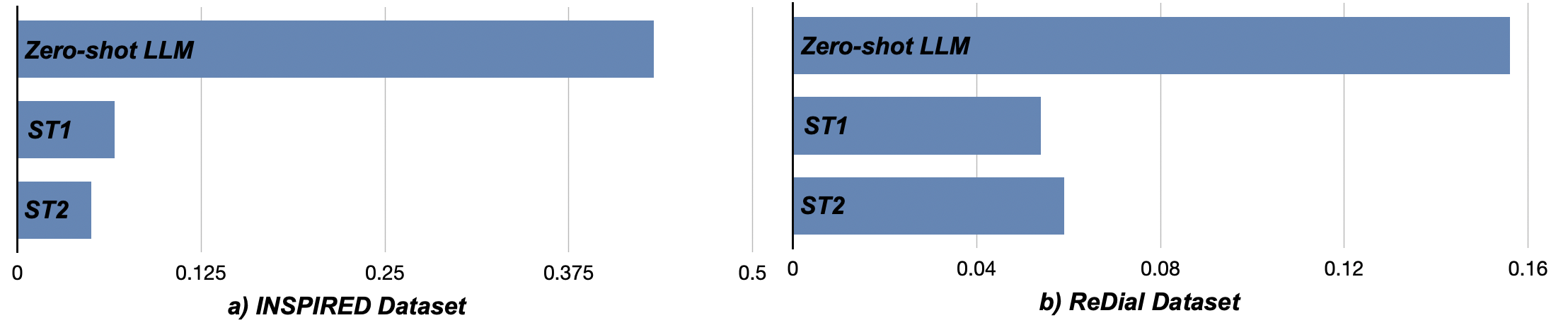}
\caption{Ratio of recommendation items out of target domain for different strategies.}
\label{OOD_figure}
\end{center}
\vspace{-3em}
\end{figure*}

\paragraph{\textbf{Adaptation Methods.}} 
In \S~\ref{Methods}, we propose 3 different adaptation methods to introduce the role of the external recommender to LLM-based CRS and 
Figure~\ref{AB_number} (left) presents a comparison of various adaptation methods across three key dimensions: \textit{a) item space information}, \textit{b) recommendation accuracy}, and \textit{c) token efficiency}. Item space information assesses how effectively each method can prevent the LLMs from generating out-of-domain items, while token efficiency reflects the token count per prompt. Recommendation accuracy compares the influence of task prompts on the final recommendation results. All metrics are normalized on a scale from 0 to 5, where higher scores indicate better outcomes. 
Our results demonstrate that both \textit{description of recommender} and \textit{self-reflection}—which incorporate detailed descriptions of external systems and their functionalities—significantly enhance recommendation accuracy and item space information compared to \textit{direct prompting}. 
However, the \textit{self-reflection} method incurs a higher token count, resulting in lower token efficiency compared to \textit{recommender description}. 
In summary, as illustrated in the radar chart (Figure~\ref{AB_number}), \textit{description of recommender} delivers the best overall performance in the CARE framework.


\paragraph{\textbf{Candidate Numbers.}}
In Figure~\ref{AB_number}, we analyze how candidate number $k$ impacts recommendation accuracy in \textit{ST2}, omitting \textit{ST1} due to its fixed parameter in $k$. 
For \textit{ST2} (default $k>20$), although the candidate number moderately affects performance (optimized $k=100$), it is relatively stable compared to the strategy selection. However, increasing the number of candidates also leads to longer input tokens, necessitating flexible parameter to optimize performance and efficiency.

\subsection{Discussion of CARE-CRS on CRS Challenges and Bias (RQ3)} \label{RQ3}
We further analyze the performance of the CARE framework in tackling key challenges and biases, such as item space discrepancy and popularity bias.


\begin{figure*}[!t]
    \centering
    \setlength{\abovecaptionskip}{5pt}   
    \setlength{\belowcaptionskip}{0pt}
    \includegraphics[width=0.9\textwidth]{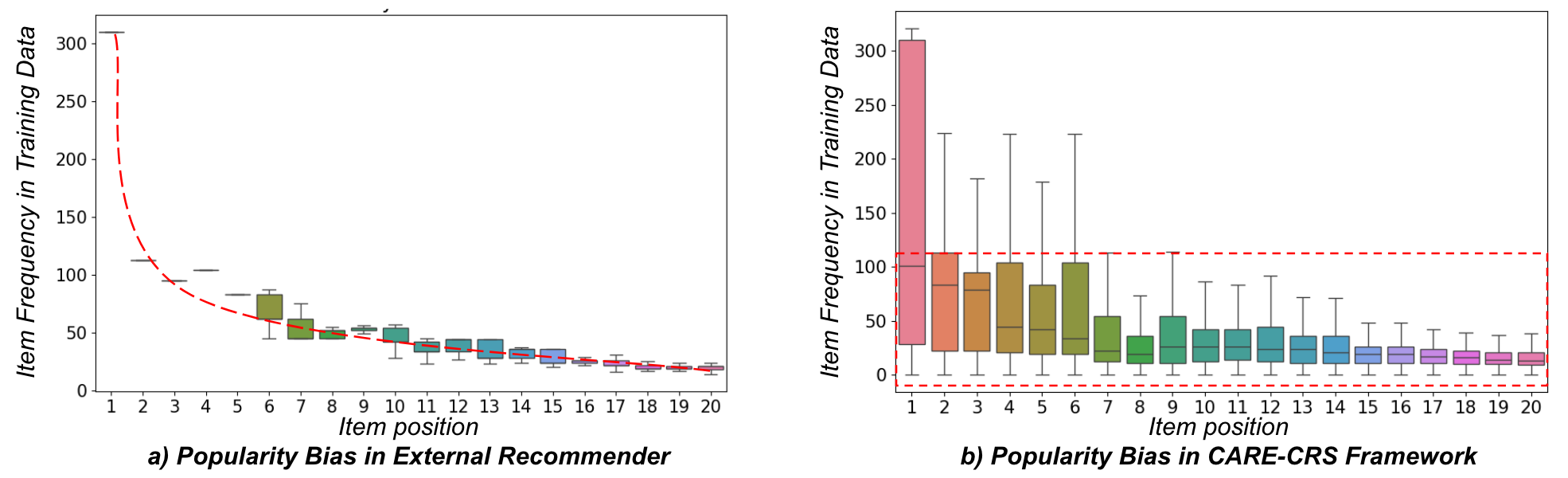}
     \caption{Popularity bias of external recommender (left) and CARE-CRS (right).}
    \label{pop_bias}
    \vspace{-1.5em}
\end{figure*}
 
\paragraph{\textbf{Item Space Discrepancy from LLMs.}} 
In Figure~\ref{OOD_figure}, we report the ratio of recommended items falling outside the target domain for each method. Because LLM-based recommendations are generated in free text, we map each recommended item to a target-domain item via Levenshtein distance, which calculates the semantic similarity between two terms; any item whose distance exceeds a threshold $\vartheta = 2 $ is classified as out-of-domain. Zero-shot LLMs often generate items that do not exist in the target domain (\emph{item space discrepancy}, \S\ref{OOD}), whereas external recommenders, which are trained to rank items solely within the target domain, exhibit no such discrepancy.
Our proposed strategies mitigate this issue by applying policy to a given candidate set. As shown in Figure~\ref{OOD_figure}, CARE reduces out-of-domain outputs and consequently improve recommendation accuracy (Figure~\ref{AB_strategy}). 


\paragraph{\textbf{Popularity Bias from Recommender.}}
As discussed in \S~\ref{contextual_engag}, popularity bias remains a significant challenge in learning-based recommendation systems \cite{pop_bias,pop_bias1}. Entity-level recommenders often reinforce this bias by consistently ranking the same popular items at the top. Figure~\ref{pop_bias} presents a box plot of item popularity to their ranking position, showing that external recommenders (with high mean and low variance) consistently rank the same popular items at the top positions. In contrast, with contextual engagement in the CARE framework, the model
explicitly understands the user's diverse intentions and preferences  (with low mean and high variance). As a result, they can either select lower-ranked items and rerank them to higher positions to form the final ranking list.
As a real example shown in Figure~\ref{overall}, even though the user prefers horror movies, the external recommender still ranks \textit{Monna} (a fantasy movie) first due to its popularity in training data, which is irrelevant to the context. However, with our method, the LLM can remove this popular item by excluding \textit{Monna} from the final ranking list. Consequently, it reranks other relevant movies such as \textit{The Exorcist} (horror) and \textit{Disturbia} (thriller) by elevating them to top positions, which aligns better with the user's stated interests and contextual information. 


\section{Conclusion}


This paper presents a new framework that integrates LLM-based reranking with external recommenders, aiming to address two key limitations of LLM-based CRS.
We employ a transformer-based recommender to perform entity-level sequential modelling and generate a candidate set of items. Subsequently, LLMs are guided to select and rerank these candidates based on conversational context through a contextual adaptation process. This design enables seamless and training-efficient collaboration between LLMs and external recommenders.
Results and analysis demonstrate the efficacy over both open- and closed-source LLMs, while mitigating the identified limitations and bias.

\subsubsection{\ackname} 

This research was supported by the Singapore Ministry of Education (MOE) Academic Research Fund (AcRF) Tier 1 grant (Proposal ID: 24-SIS-SMU-002) and the Lee Kong Chian Fellowship awarded to DENG Yang by Singapore Management University. 

\subsubsection{\discintname}
 The authors have no competing interests to declare that are relevant to the content of this article. 
%
%
%
\bibliographystyle{splncs04}
\bibliography{CLEAR}

%

\end{document}